\documentclass[12pt]{article} 
\usepackage{amsmath}
\usepackage{amssymb}
\usepackage{axodraw}
\usepackage{epsfig}
\usepackage{mcite}

\newcommand{\be}{\begin{equation}}
\newcommand{\ee}{\end{equation}}
\newcommand{\ba}{\begin{eqnarray}}
\newcommand{\ea}{\end{eqnarray}}

\newcommand{\gev    }{\ensuremath{\mathrm{GeV}}}
\newcommand{\gevsq  }{\ensuremath{\mathrm{GeV^2}}}
\newcommand{\Lumi}{\ensuremath{{\cal L}}}

\newcommand{\gam}{\ensuremath{(\Delta)\gamma^N}}
\newcommand{\gamel}{\ensuremath{(\Delta)\gamma^N_{\rm el}}}
\newcommand{\gaminel}{\ensuremath{(\Delta)\gamma^N_{\rm inel}}}
\newcommand{\gamelp}{\ensuremath{\gamma^p_{\rm el}}}
\newcommand{\gaminelp}{\ensuremath{\gamma^p_{\rm inel}}}
\newcommand{\der}{\ensuremath{{\operatorname{d}}}}
\newcommand{\av}[1]{\ensuremath{\langle{#1}\rangle}}

\newcommand{\shat}{\ensuremath{\hat{s}}}
\newcommand{\that}{\ensuremath{\hat{t}}}
\newcommand{\uhat}{\ensuremath{\hat{u}}}
\newcommand{\vmin}[1]{\ensuremath{#1_{\rm min}}}
\newcommand{\vmax}[1]{\ensuremath{#1_{\rm max}}}

\begin{document} 
\setlength{\parskip}{0.45cm} 
\setlength{\baselineskip}{0.75cm} 
\numberwithin{equation}{section}

%
%
%
\begin{titlepage} 
\setlength{\parskip}{0.25cm} 
\setlength{\baselineskip}{0.25cm} 
\begin{flushright} 
DO-TH 2002/16\\ 
\vspace{0.2cm} 
hep--ph/0209335\\ 
\vspace{0.2cm} 
September 2002 
\end{flushright} 
\vspace{1.0cm} 
\begin{center} 
\LARGE 
{\bf Delineating the polarized and unpolarized photon
distributions of the nucleon in $eN$ collisions} 
\vspace{1.5cm} 
 
\large 
M. Gl\"uck, C.\ Pisano, E.\ Reya, I.\ Schienbein\\
\vspace{1.0cm} 
 
\normalsize 
{\it Universit\"{a}t Dortmund, Institut f\"{u}r Physik,}\\ 
{\it D-44221 Dortmund, Germany} \\ 
\vspace{0.5cm}

\vspace{1.5cm} 
\end{center} 
 
\begin{abstract} 
\noindent 
The production rates of lepton-photon and dimuon pairs
at the HERA collider and the HERMES experiment are evaluated
in the leading order equivalent photon approximation.
It is shown that the production rates are sufficient to 
facilitate the extraction of the polarized and unpolarized
equivalent photon distributions of the proton and neutron
in the available kinematical regions. It is pointed out that
these results indicate the possibility of additional, independent,
tests concerning the  unpolarized and polarized structure functions
$F_{1,2}^N$ and $g_{1,2}^N$, respectively, of the nucleon.
\end{abstract} 
\end{titlepage} 
 
\section{Introduction}
In a previous publication \cite{Gluck:2002fi} we presented
the polarized and unpolarized equivalent photon distributions
$\gam(y,Q^2)$ of the nucleon, $N=p,n$, consisting of two
components,
\be
\gam(y,Q^2) = \gamel(y,Q^2) + \gaminel(y,Q^2)
\label{eq:gam}
\ee
where the elastic parts, $\gamel$, are due to 
$N \to \gamma N$ while the inelastic parts, $\gaminel$,
derive from $N \to \gamma X$ with $X \ne N$.
It turns out that, as in the case of $\gamelp(y,Q^2)$ studied
in \cite{Kniehl:1991iv}, $\gamel(y,Q^2)$ are uniquely determined by the
well known electromagnetic form factors $F^N_{1,2}(q^2)$ of the 
nucleon. 
The inelastic components were fixed via the boundary conditions
\cite{Gluck:2002fi}
\be
\gaminel(y,Q_0^2) = 0
\label{eq:bc}
\ee
at $Q_0^2 = 0.26\ \gevsq$, evolved for $Q^2 > Q_0^2$ according to
the leading order (LO) equation
\be
\frac{\der \gaminel(y,Q^2)}{\der \ln Q^2} =
\frac{\alpha}{2 \pi} \sum_{q=u,d,s} e_q^2 \int_y^1 \frac{\der x}{x}\ 
(\Delta) P_{\gamma q}\left(\frac{y}{x}\right)
\left[(\Delta) q^N(x,Q^2) + (\Delta) \bar{q}^N(x,Q^2)\right]
\label{eq:evolution}
\ee
with the unpolarized and polarized parton distributions in
LO taken from \cite{Gluck:1998xa} and \cite{Gluck:2000dy}.

As stated in \cite{Gluck:2002fi}, the boundary conditions
\eqref{eq:bc} are not compelling but should be tested experimentally.
However at large scales $Q^2$ the results
become rather insensitive to details at the input scale $Q_0^2$ and
thus the vanishing boundary conditions \eqref{eq:bc} yield reasonable
results for $\gaminel$ which are essentially determined by the quark
and antiquark (sea) distributions of the nucleon in \eqref{eq:evolution}.
At low scales $Q^2$, however, $\gaminel(y,Q^2)$ depend obviously
on the assumed details at the input scale $Q_0^2$.
Such a situation is encountered at a fixed target experiment, typically
HERMES at DESY. 
At present it would be too speculative and arbitrary
to study the effects due to a non-vanishing boundary 
$\gaminel(y,Q_0^2) \ne 0$.
Rather this should be examined experimentally if our expectations 
based on the  vanishing boundary \eqref{eq:bc} turn out to be
in disagreement with observations.

The photon distributions $\gam$ of the nucleon, being the counterparts
of the well known photon distribution of the electron
$\gamma^e(y,Q^2)$, are useful for cross section estimates in the
equivalent photon approximation which simplifies more involved exact
calculations (\cite{Courau:1992ht}, for example).
Thus measurements of $\gam(y,Q^2)$ are not only interesting on their own,
but may provide additional information concerning 
$(\Delta) \overset{(-)}{q}$$^N$ 
in \eqref{eq:evolution}, in particular
about the polarized parton distributions which are not well determined
at present.

In the present paper we consider muon pair production
$e N \to e \mu^+ \mu^- X$ via the subprocess 
$\gamma^e \gamma^N \to \mu^+ \mu^-$ and the Compton process
$e N \to e \gamma X$ via the subprocess 
$e \gamma^N \to e \gamma$ for both the HERA collider experiments
and the polarized and unpolarized fixed target HERMES experiment at
DESY.
The Compton scattering process at HERA has already been studied in 
the equivalent photon approximation \cite{DeRujula:1998yq} as well as
in an exact calculation \cite{Courau:1992ht}. 
It should be noted that a study of 
$N N \to \mu^+ \mu^- X$ via 
$\gamma^N \gamma^N \to \mu^+ \mu^-$
in hadron-hadron collisions is impossible 
\cite{Drees:1994zx,*Ohnemus:1994qw}
due to the dominance of the Drell-Yan subprocess 
$q^N \bar{q}^N \to \mu^+ \mu^-$.
The measurements at HERMES provide the unique opportunity of getting
information concerning the {\underline{polarized}} photon distributions,
$\Delta \gamma^N$, of the nucleon as well.

Our equivalent photon event rate estimates provide furthermore information
concerning the possibility of measuring the polarized structure
functions $g_1^N$ via Compton scattering or dimuon production along
the lines of \cite{Courau:1992ht} as extended to the 
spin dependent situation \cite{winp}.

\section{Theoretical Framework}
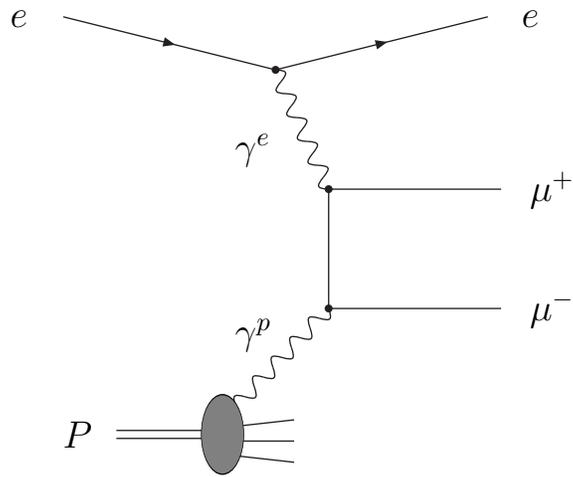
\begin{figure}[ht]
\begin{center}
\begin{picture}(210,200)(0,0)
\ArrowLine(20,180)(100,160)
\ArrowLine(100,160)(180,180)
\Vertex(100,160){1.5}
\Photon(100,160)(120,115){3}{5}
\Photon(85,33)(120,70){3}{5}
\Line(40,21)(80,21)
\Line(40,24)(80,24)
\Line(80,25)(107,28)
\Line(80,20)(107,20)
\Line(80,15)(107,12)
\GOval(80,22.5)(15,8)(0){0.5}
\Line(185,70)(120,70)
\Line(120,70)(120,115)
\Line(120,115)(185,115)
\Vertex(120,70){1.5}
\Vertex(120,115){1.5}
%
\Text(0,180)[cl]{\large $e$}
\Text(200,180)[cr]{\large $e$}
\Text(205,115)[c]{\large $\mu^+$}
\Text(205,70)[c]{\large $\mu^-$}
\Text(20,22.5)[cl]{\large $P$}
\Text(85,60)[l]{\large $\gamma^p$}
\Text(85,130)[l]{\large $\gamma^e$}
\end{picture}

\end{center}
\caption{\sf Lowest-order Feynman diagram for dimuon production
in $ep$ collisions. (The crossed $\uhat$-channel diagram
is not shown).}
\label{fig:dimuon}
\end{figure}

Considering first deep inelastic dimuon production
$e p \to e \mu^+ \mu^- X$ at HERA ($s=4 E_e E_p$) via the subprocess
$\gamma^e \gamma^p \to \mu^+ \mu^-$ as depicted in 
Fig.~\ref{fig:dimuon}, let $\eta_1$ and $\eta_2$ be the
(laboratory-frame) rapidities of $\mu^+$ and $\mu^-$
measured along the proton beam direction.
Then the production process can be written as\footnote{A useful summary
of the relevant kinematics can be found in Appendix D of
\protect\cite{Brock:1995sz} where the c.m. rapidities $y_i$ have been
used which are related to our laboratory-frame rapidities $\eta_i$
via $y_i = \eta_i - \ln \sqrt{E_p/E_e}$ for HERA ($\eta_i$ is defined
to be positive in the proton forward direction) and 
$y_i = \eta_i + \ln \sqrt{M/2 E_e}$ for HERMES 
(with $\eta_i$ being positive in the electron forward direction).
Notice that, besides $\eta_1$ and $\eta_2$, we have chosen $\xi$ in
\protect\eqref{eq:xi} as third independent kinematical variable in
\protect\eqref{eq:dimuon-hera} instead of the more commonly used
$\shat$ or the transverse momentum $p_T$ of one of the two muons
(which balance each other in LO), related by 
$\shat = 2 p_T^2 [1+\cosh(\eta_1-\eta_2)]$.}
\be
\frac{\der \sigma}{\der \eta_1 \der \eta_2 \der \xi}
= \frac{4 \xi E_e^2}{1+\cosh(\eta_1-\eta_2)}
\frac{e^{\eta_1}+e^{\eta_2}}{e^{-\eta_1}+e^{-\eta_2}}
\xi \gamma^e(\xi,\shat) x \gamma^p(x,\shat)
\frac{\der \hat \sigma}{\der \that}
\label{eq:dimuon-hera}
\ee
where $\shat = (p_{\mu^+}+p_{\mu^-})^2$ denotes the dimuon invariant
mass squared and the measured four-momenta
$p_{\mu^+,\mu^-}$ of the produced muons fix
the momentum fractions either via
\ba
\xi &=& \frac{\sqrt{\shat}}{2 E_e} 
\left(\frac{e^{-\eta_1}+e^{-\eta_2}}{e^{\eta_1}+e^{\eta_2}} \right)^{1/2}
\label{eq:xi}
\\
x &=& \frac{\sqrt{\shat}}{2 E_p} 
\left(\frac{e^{\eta_1}+e^{\eta_2}}{e^{-\eta_1}+e^{-\eta_2}} \right)^{1/2}
\label{eq:x}
\ea
or equivalently via $x E_p + \xi E_e = p_{\mu^+}^0+p_{\mu^-}^0$ and
$4\xi x E_e E_p = \shat$ where
$E_p = 820\ \gev$, $E_e = 27.5\ \gev$ are the colliding proton
and electron energies.
In the spirit of the leading order equivalent photon
approximation underlying \eqref{eq:dimuon-hera}, we shall adopt
the LO photon distribution $\gamma^p(x,\shat)$ of the proton
in \cite{Gluck:2002fi} as well as the LO equivalent photon distribution
$\gamma^e(\xi,\shat)$ of the electron,
\be
\gamma^e(\xi,\shat) = \frac{\alpha}{2 \pi} \frac{1+(1-\xi)^2}{\xi}
\ln \frac{\shat}{m_e^2}\ .
\label{eq:gamma_e}
\ee
The cross section $\frac{\der \hat \sigma}{\der \that}$
in \eqref{eq:dimuon-hera} for the subprocess 
$\gamma^e \gamma^p \to \mu^+ \mu^-$ reads
\be
\frac{\der \hat \sigma}{\der \that}^{\gamma \gamma \to \mu^+ \mu^-} 
= \frac{2 \pi \alpha^2}{\shat^2}
\left(\frac{\that}{\uhat}+ \frac{\uhat}{\that}\right)
= \frac{4 \pi \alpha^2}{\shat^2} \cosh(\eta_1 - \eta_2)\ .
\label{eq:sub-dimuon}
\ee

\begin{figure}[h]
\begin{center}
\begin{picture}(240,100)(10,0)
\ArrowLine(60,90)(120,70)
\ArrowLine(120,70)(160,70)
\ArrowLine(160,70)(220,90)
\Vertex(120,70){1.5}
\Photon(160,70)(210,40){3}{5}
\Vertex(160,70){1.5}
\Photon(85,33)(120,70){3}{5}
\Line(40,21)(80,21)
\Line(40,24)(80,24)
\Line(80,25)(107,28)
\Line(80,20)(107,20)
\Line(80,15)(107,12)
\GOval(80,22.5)(15,8)(0){0.5}
%
\Text(40,90)[cl]{\large $e$}
\Text(240,90)[cr]{\large $e$}
\Text(230,40)[cr]{\large $\gamma$}
\Text(20,22.5)[cl]{\large $P$}
\Text(83,58)[l]{\large $\gamma^p$}
\end{picture}

\end{center}
\caption{\sf Lowest-order Feynman diagram for Compton scattering
in $ep$ collisions. (The crossed $\uhat$-channel contribution
is not shown).}
\label{fig:compton}
\end{figure}
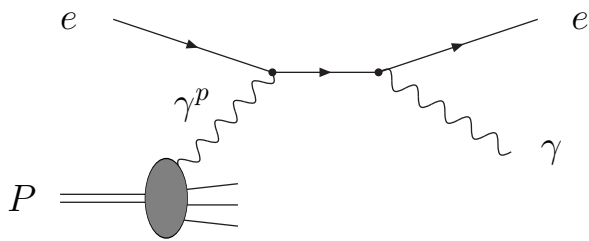

For the Compton process $e p \to e \gamma X$, proceeding via the
subprocess $e \gamma^p \to e \gamma$ as depicted in Fig.~\ref{fig:compton},
Eq.~\eqref{eq:dimuon-hera} is replaced by
\be
\frac{\der \sigma}{\der \eta_e \der \eta_\gamma}
= \frac{4 E_e^2}{1+\cosh(\eta_e-\eta_\gamma)}
\frac{e^{\eta_e}+e^{\eta_\gamma}}{e^{-\eta_e}+e^{-\eta_\gamma}}
x \gamma^p(x,\shat)
\frac{\der \hat \sigma}{\der \that}\ ,
\label{eq:dics-hera}
\ee
i.e., $\xi =1$ and with $\eta_{e,\gamma}$ the rapidities of the
produced (outgoing) electron and photon measured, again, in the
proton beam direction.
The cross section $\der \hat\sigma/\der \that$ for the subprocess
$e \gamma^p \to e \gamma$ reads
\be
\frac{\der \hat \sigma}{\der \that}^{e \gamma \to e \gamma} 
= - \frac{2 \pi \alpha^2}{\shat^2}
\left(\frac{\shat}{\uhat}+ \frac{\uhat}{\shat}\right)
\label{eq:sub-compton}
\ee
with $\shat =(p_e + p_\gamma)^2$ and 
$-\shat/\hat{u}=1+e^{\eta_e-\eta_\gamma}$. Here $x$ is fixed
by \eqref{eq:x} or by either $E_e + x E_p = p_e^0 + p_\gamma^0$
or $4 x E_e E_p = \shat$.

The extension to the fixed-target experiment HERMES ($s=2 M E_e$)
is obtained via $E_p \to M/2$ and $\eta_i \to - \eta_i$
everywhere with $\eta_i$ now corresponding to the rapidities of
the observed particles with respect to the electron beam direction.
Furthermore, at HERMES one may study also $\gamma^n(x,\shat)$
as well as the polarized $\Delta \gamma^N(x,\shat)$ in 
\cite{Gluck:2002fi} by utilizing
\be
\Delta \gamma^e(\xi,\shat) = 
\frac{\alpha}{2 \pi} \frac{1-(1-\xi)^2}{\xi}
\ln \frac{\shat}{m_e^2}
\label{eq:polgamma_e}
\ee
in the obvious spin dependent counterpart of \eqref{eq:dimuon-hera},
while the relevant LO cross sections for the polarized subprocesses
are given by
\ba
\frac{\der \Delta \hat \sigma}{\der \that}^{\gamma \gamma \to \mu^+ \mu^-} 
&=&
-\frac{\der \hat \sigma}{\der \that}^{\gamma \gamma \to \mu^+ \mu^-} 
\label{eq:sub-poldimuon}
\\
\frac{\der \Delta \hat \sigma}{\der \that}^{e \gamma \to e \gamma} 
\hspace*{5mm}
&=& - \frac{2 \pi \alpha^2}{\shat^2}
\left(\frac{\shat}{\uhat}- \frac{\uhat}{\shat}\right)
\label{eq:sub-polcompton}
\ea
with
$-\shat/\hat{u}=1+e^{\eta_\gamma-\eta_e}$. 
These expressions apply obviously also to the COMPASS $\mu p$ experiment
at CERN whose higher incoming lepton energies
($E_\mu = 50 - 200\ \gev$) enable the determination of
$\Delta \gamma^N(x,Q^2)$ at lower values of $x$ as compared to the
corresponding measurements at HERMES.
(Notice that for a muon beam one has obviously to replace
$m_e$ by $m_\mu$ in \eqref{eq:gamma_e} and \eqref{eq:polgamma_e}).

\section{Results}
We shall present here the expected number of events for the
accessible $x$-bins at HERA collider experiments and at 
the fixed target HERMES experiment subject to some 
representative kinematical cuts which, of course, may be slightly
modified in the actual experiments.
These cuts entail $\shat \ge \vmin{\shat}$, 
$\vmin{\eta} \le \eta_i \le \vmax{\eta}$ and 
$E_i \ge \vmin{E}$ where $E_i$ are the energies of the observed
outgoing particles.
The relevant integration ranges at HERA are fixed via
$0 \le \xi \le 1$, $\vmin{\shat}/4\xi E_e E_p \le x \le 1$ with
$\shat$ given by $\shat = 4 x \xi E_e E_p$ while
$\eta_i$ are constrained by $\eta_1 + \eta_2 = \ln \tfrac{x E_p}{\xi E_e}$
which follows from \eqref{eq:x}.
Here $\xi = 1$, $\eta_1 = \eta_\gamma$, $\eta_2 = \eta_e$ for
the Compton scattering process, Eq.~\eqref{eq:dics-hera}.
The relation 
$\eta_i - \eta_j = \ln[\tfrac{\xi E_e}{E_i}(1+ e^{2 \eta_i})-1]$
as obtained from the outgoing particle energy $E_i$ and its transverse
momentum \cite{Brock:1995sz}, further restricts the integration range of
$\eta_{i,j}$ as dictated by $E_i \ge \vmin{(E_i)}$.
At HERMES $E_p \to M/2$ and $\eta_i \to - \eta_i$ in the above 
expressions with $\eta_i$ the outgoing particle rapidity
with respect to the ingoing lepton direction.

In the following we shall consider $\vmin{E} =4\ \gev$.
For the Compton scattering process we further employ
$\vmin{\shat} = 1\ \gevsq$ so as to guarantee the applicability of 
perturbative QCD, i.e., the relevance of the utilized \cite{Gluck:2002fi}
$\gam(x,\shat)$.
For the dimuon production process we shall impose 
$\vmin{\shat} = M^2[\Psi(2 S)] = (3.7\ \gev)^2$ 
so as to evade the dimuon background
induced by charmonium decays at HERMES (higher charmonium states have 
negligible branching ratios into dimuons);
for HERA we impose in addition 
$\vmax{\shat} = M^2[\Upsilon(1S)]= (9.4\ \gev)^2$ in order to avoid
the dimuon events induced by bottonium decays.
Finally, at HERA we consider $\vmin{\eta}=-3.8$, $\vmax{\eta}=3.8$
and at HERMES $\vmin{\eta}= 2.3$, $\vmax{\eta}=3.9$.
The integrated luminosities considered are
$\Lumi_{\rm HERA} = 100\ pb^{-1}$ and
$\Lumi_{\rm HERMES} = 1\ fb^{-1}$.

\begin{figure}[t]
\centering
\epsfig{figure=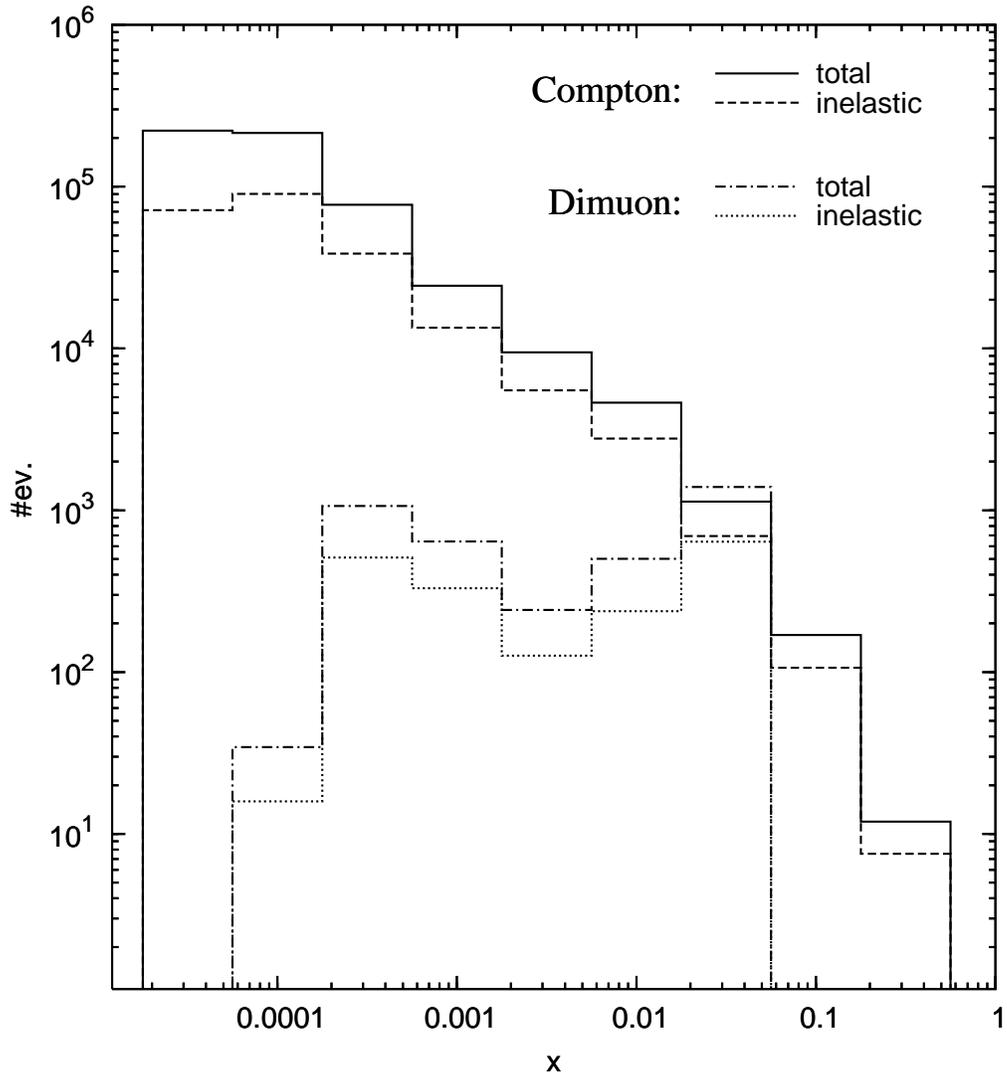, width = 14cm}

\vspace*{1cm}
\caption{
Event rates for Compton ($e\gamma \to e \gamma$) and dimuon 
production ($\gamma \gamma \to \mu^+ \mu^-$) processes at the 
HERA collider. The cuts applied are as described in the text.}
\label{fig:fig3}
\end{figure}

\begin{figure}[t]
\centering
\epsfig{figure=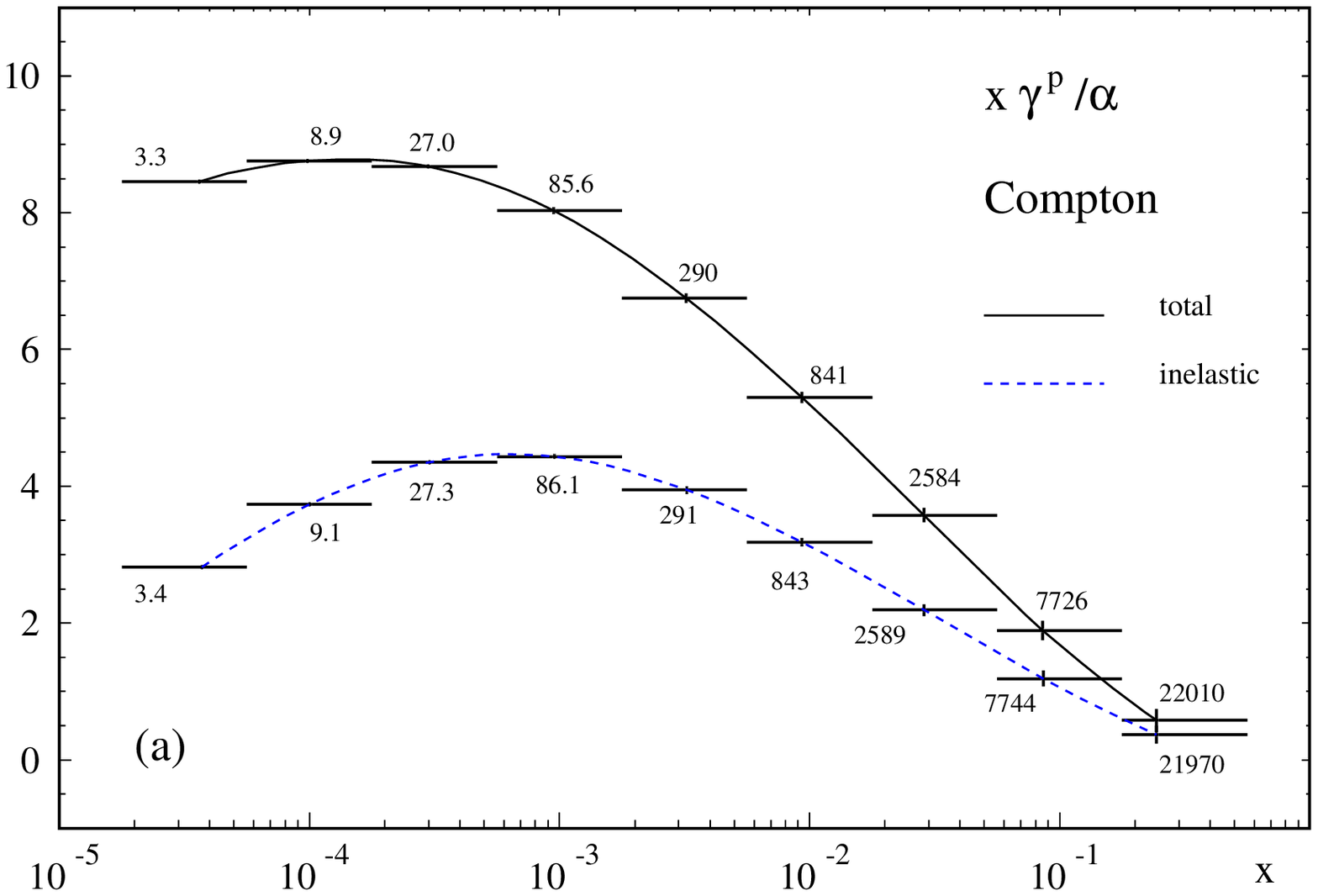, width = 14cm}
\epsfig{figure=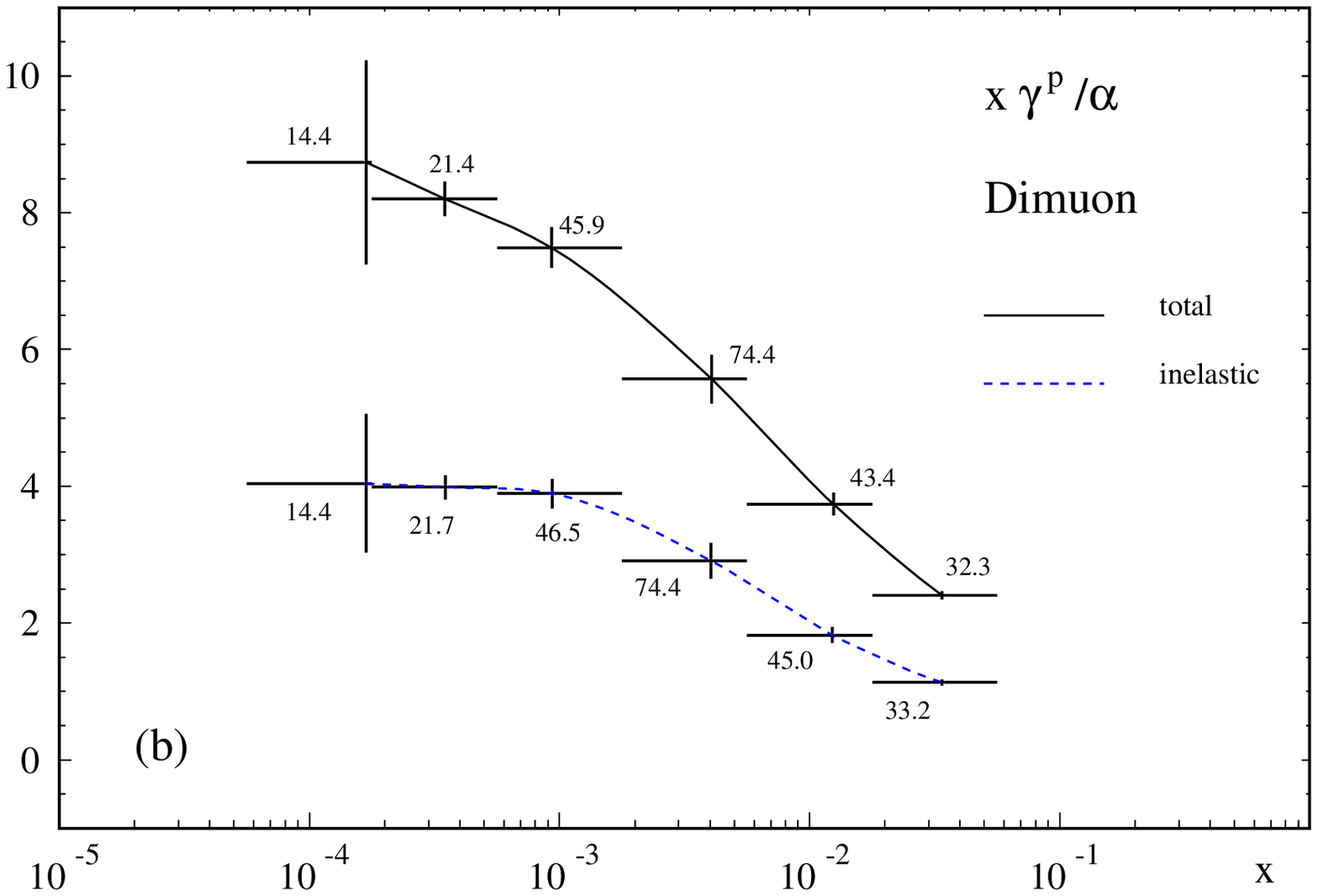, width = 14cm}

\vspace*{1cm}
\caption{
Expected statistical accuracy of the determination
of $\gamma^p(\av{x},\av{\shat})$ via the (a) Compton process
and (b) the dimuon production process at the HERA collider.
The numbers indicate the average scale $\av{\shat}$ (in $\gevsq$ units)
for each $x$-bin.}
\label{fig:fig4}
\end{figure}

\begin{figure}[t]
\centering
\epsfig{figure=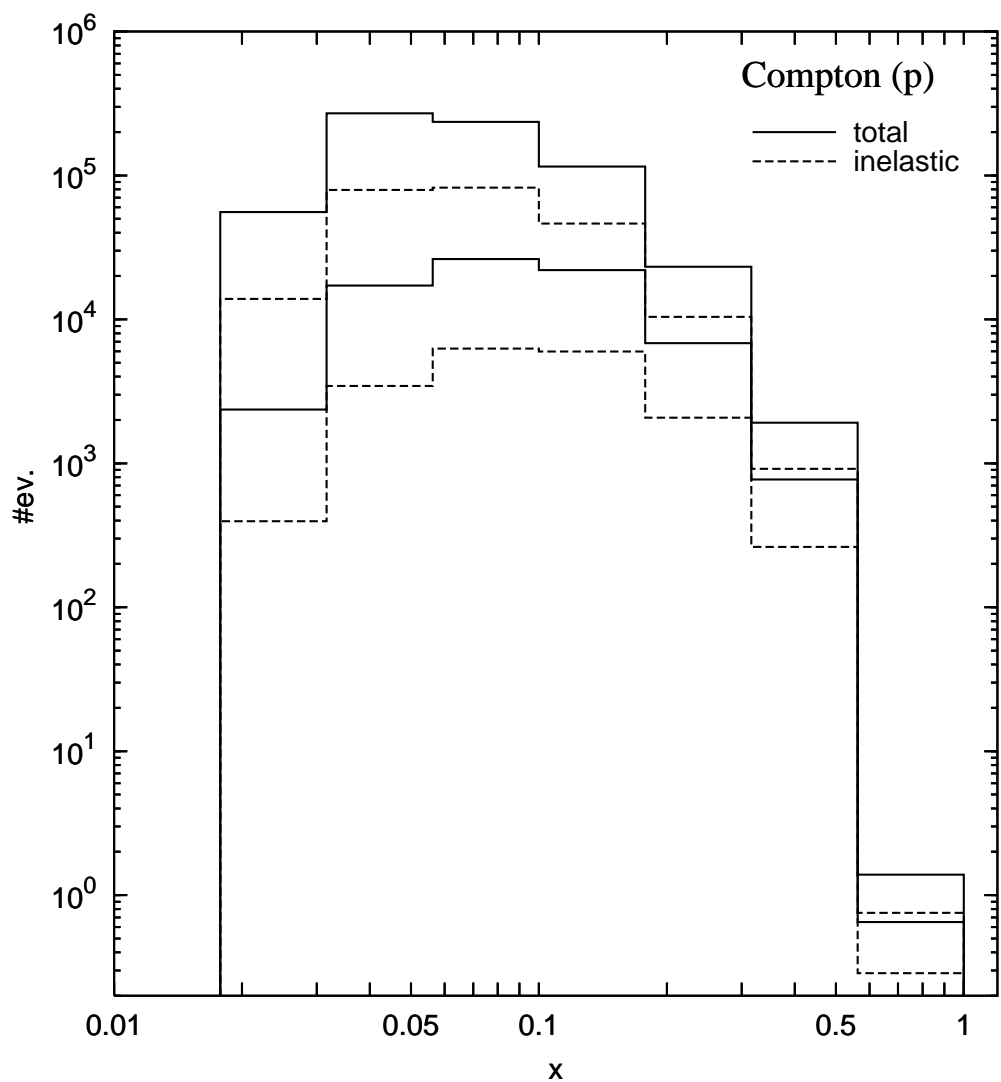, width = 14cm}

\vspace*{1cm}
\caption{
Event rates for the Compton process at HERMES using an
(un)polarized proton target. The upper (solid and dashed)
curves refer to an unpolarized proton, whereas the lower ones
refer to a polarized proton target. The cuts applied are as described
in the text.}
\label{fig:fig5}
\end{figure}

In Fig.~\ref{fig:fig3} the histograms depict the expected number
of dimuon and Compton events at HERA found by integrating 
Eqs.~\eqref{eq:dimuon-hera} and \eqref{eq:dics-hera} 
applying the aforementioned cuts 
and constraints.
The important inelastic contribution due to $\gaminelp$ in 
\eqref{eq:gam}, being calculated according to \eqref{eq:evolution}
using the minimal boundary condition \eqref{eq:bc}, is shown
separately by the dashed curves.
To illustrate the experimental extraction of $\gamma^p(x,\shat)$ we
translate the information in Fig.~\ref{fig:fig3} into a statement
on the accuracy of a possible measurement by evaluating 
$\gamma^p(\av{x},\av{\shat})$ at the averages
$\av{x}$, $\av{\shat}$ determined from the event sample in
Fig.~\ref{fig:fig3}.
Assuming that in each bin the error is only statistical, i.e.
$\delta \gamma = \pm \gamma/ \sqrt{N_{\rm bin}}$,
the results for $x \gamma/\alpha$ are shown in
Fig.~\ref{fig:fig4}. It should be noticed that the statistical
accuracy shown will increase if $\gaminelp(x,Q_0^2) \ne 0$ in
contrast to our vanishing boundary condition \eqref{eq:bc} used
in all our present calculations. Our results for the Compton process
in Figs.~\ref{fig:fig4} and~\ref{fig:fig5} are, apart from our somewhat
different cut requirements, similar to the ones presented in 
{\cite{DeRujula:1998yq}}.

\begin{figure}[t]
\centering
\epsfig{figure=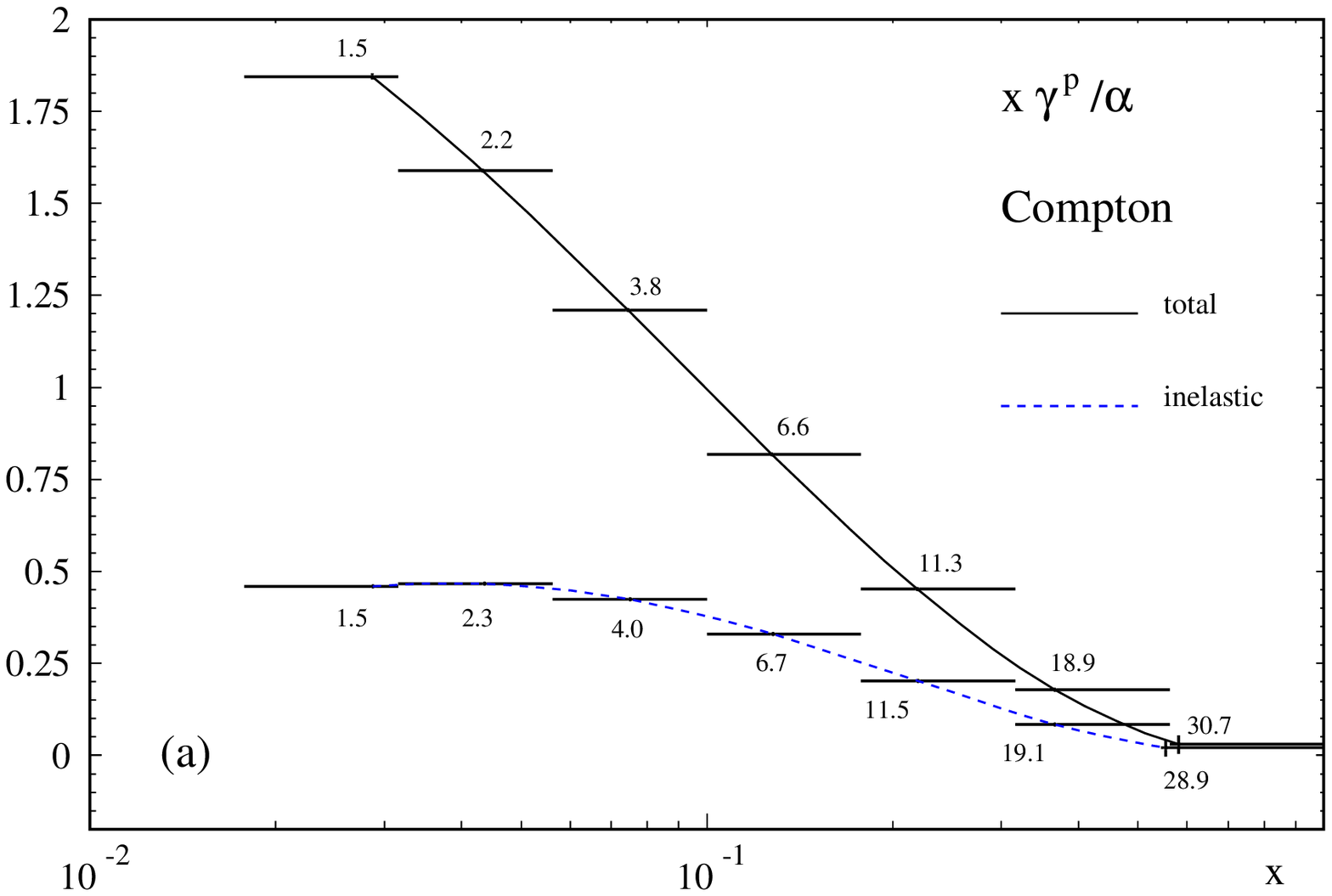, width = 14cm}
\epsfig{figure=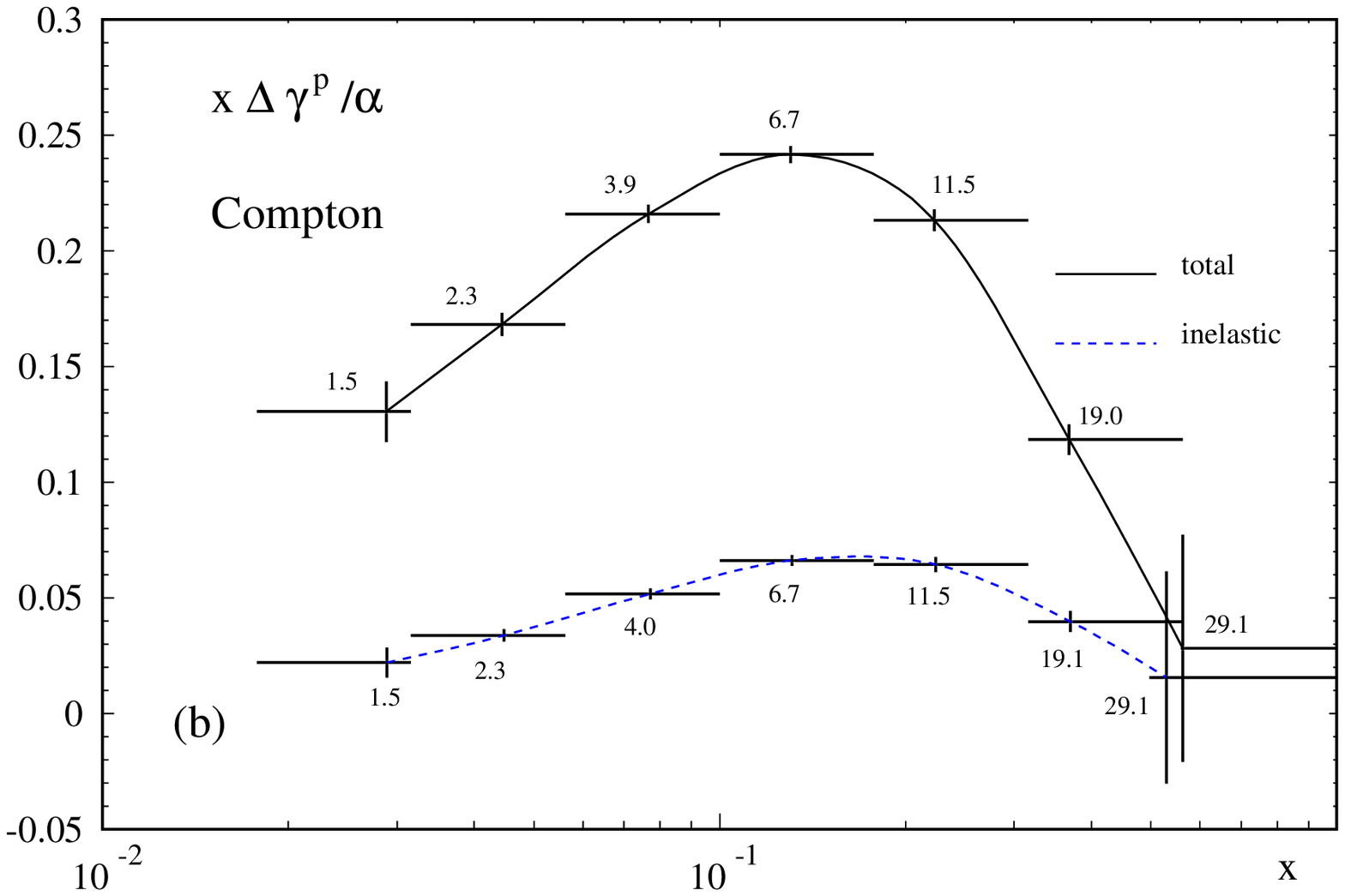, width = 14cm}

\vspace*{1cm}
\caption{
Expected statistical accuracy of the determination
of (a) $\gamma^p(\av{x},\av{\shat})$ 
and (b) $\Delta \gamma^p(\av{x},\av{\shat})$ 
via the Compton process at HERMES using an (un)polarized
proton target.
The numbers indicate the average scale $\av{\shat}$ 
(in $\gevsq$ units) for each bin.}
\label{fig:fig6}
\end{figure}

\begin{figure}[t]
\centering
\epsfig{figure=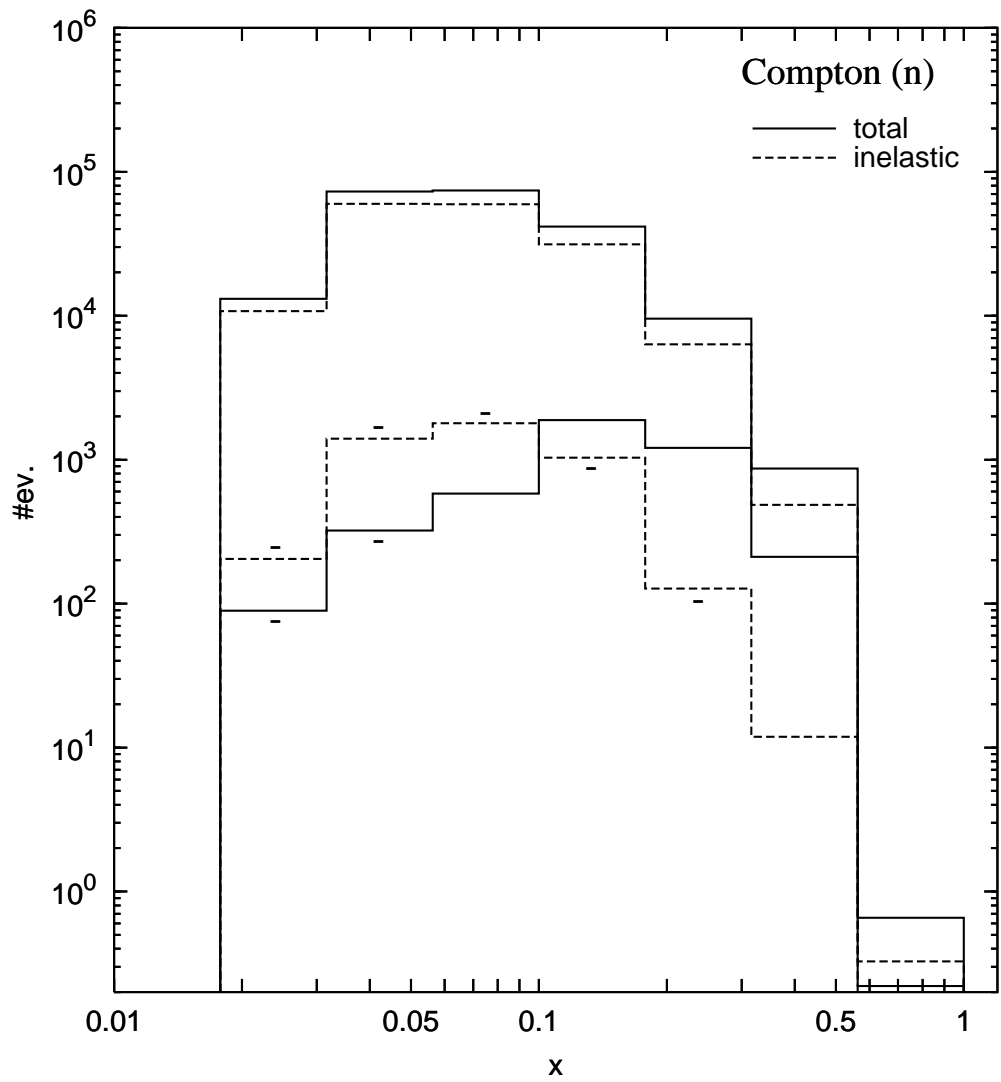, width = 14cm}

\vspace*{1cm}
\caption{
As in Fig.~\protect\ref{fig:fig5} but for a neutron target.
The negative signs at some lower-$x$ bins indicate that the 
polarized total cross section and/or inelastic contribution
is negative.}
\label{fig:fig7}
\end{figure}

\begin{figure}[t]
\centering
\epsfig{figure=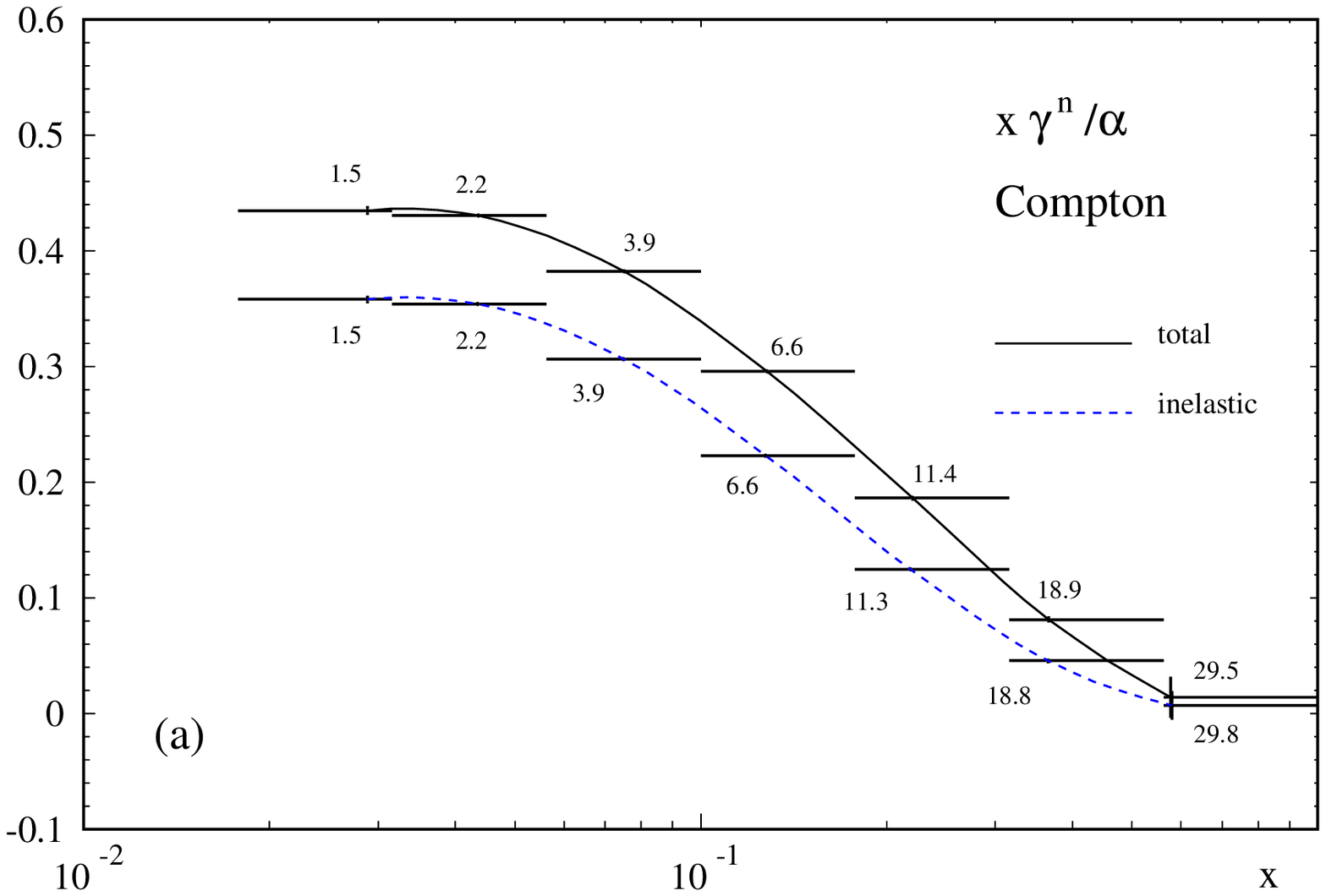, width = 14cm}
\epsfig{figure=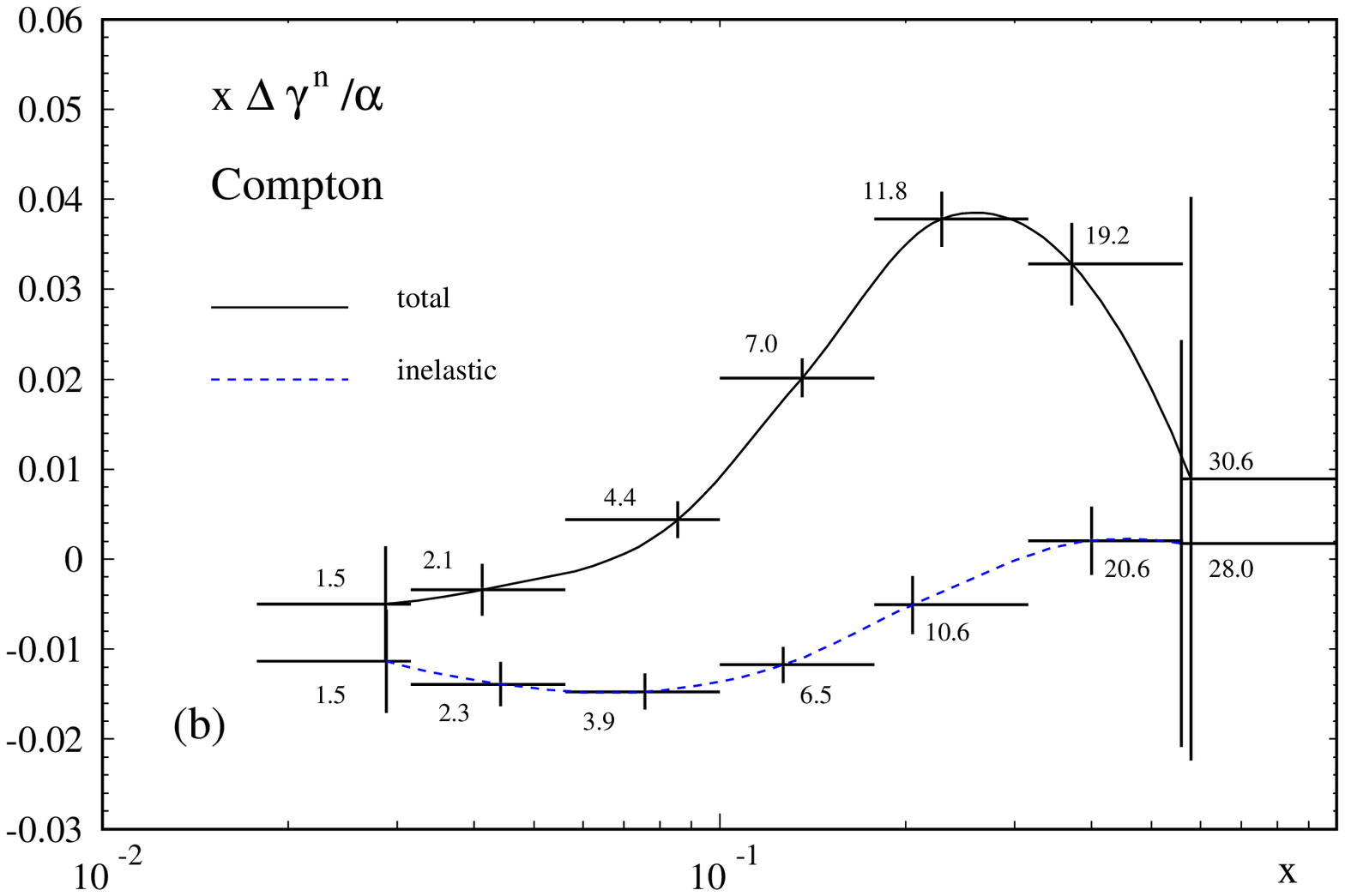, width = 14cm}

\vspace*{1cm}
\caption{
As in Fig.~\protect\ref{fig:fig6} but for a neutron target.}
\label{fig:fig8}
\end{figure}

Apart from testing $\gamma^N(x,\shat)$ at larger values of $x$, the
fixed-target HERMES experiment can measure the polarized
$\Delta \gamma^N(x,\shat)$ as well.
In Fig.~\ref{fig:fig5} we show the expected number of Compton events
for an (un)polarized proton target. The accuracy of a possible 
measurement of $\gamma^p(\av{x},\av{\shat})$ and
$\Delta \gamma^p(\av{x},\av{\shat})$ is illustrated in 
Fig.~\ref{fig:fig6} where the averages
$\av{x}$, $\av{\shat}$ are determined from the event sample in
Fig.~\ref{fig:fig5} by assuming that the error is only statistical
also for the polarized photon distribution, i.e.
$\delta(\Delta \gamma) = \pm (\sqrt{N_{\rm bin}}/|\Delta N_{\rm bin}|)
\Delta \gamma$.
The analogous expectations for an (un)polarized neutron target are
shown in Figs.~\ref{fig:fig7} and \ref{fig:fig8}.
It should be pointed out that, according to Figs.~\ref{fig:fig6}(b)
and \ref{fig:fig8}(b), HERMES measurements will be sufficiently
accurate to delineate even the polarized $\Delta \gamma^{p,n}$ 
distributions in the medium- to small-x region, in particular the
theoretically more speculative inelastic contributions.

\begin{figure}[t]
\vspace*{-1.0cm}
\centering
\epsfig{figure=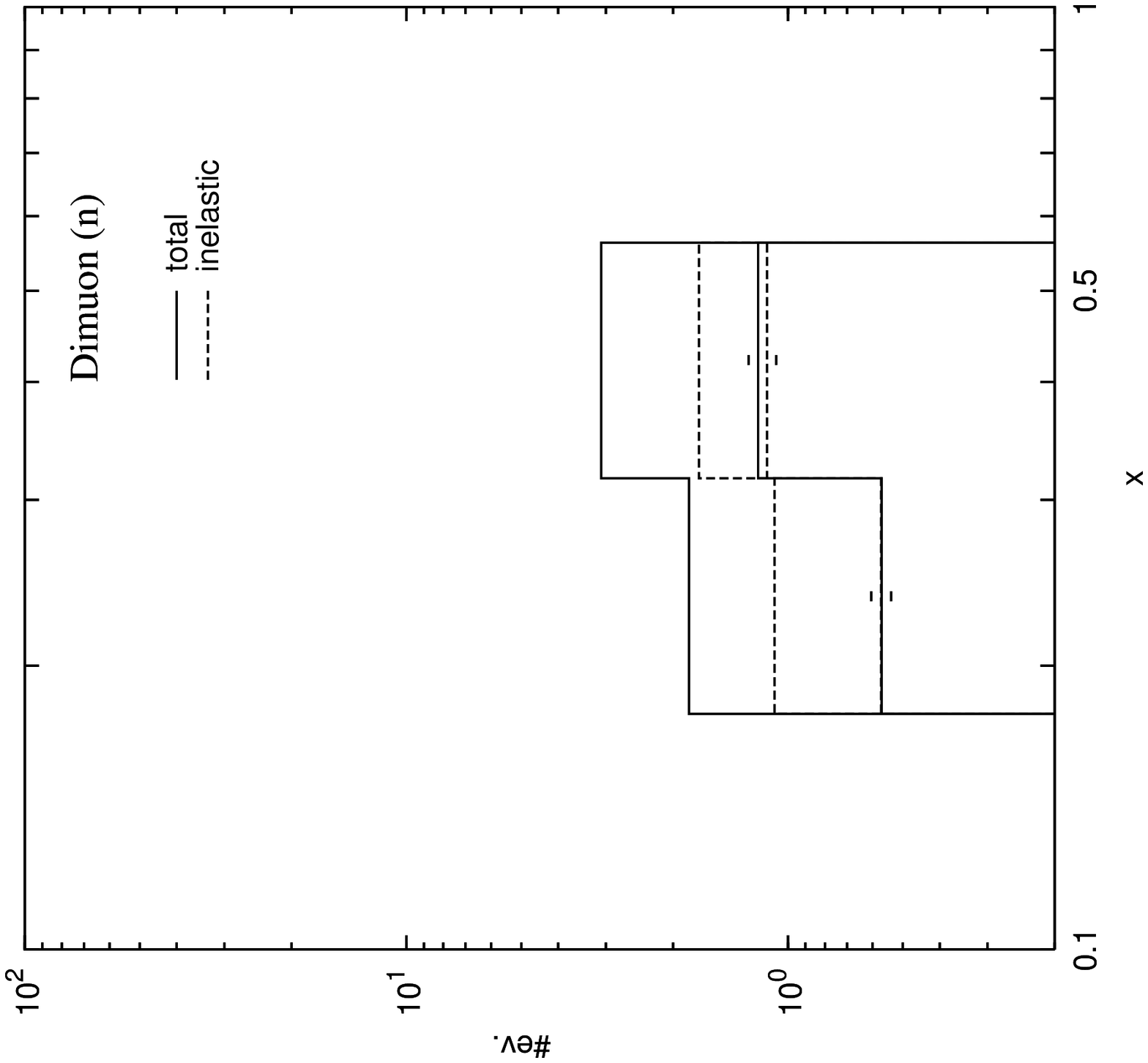, angle = 90, width = 11cm}
\epsfig{figure=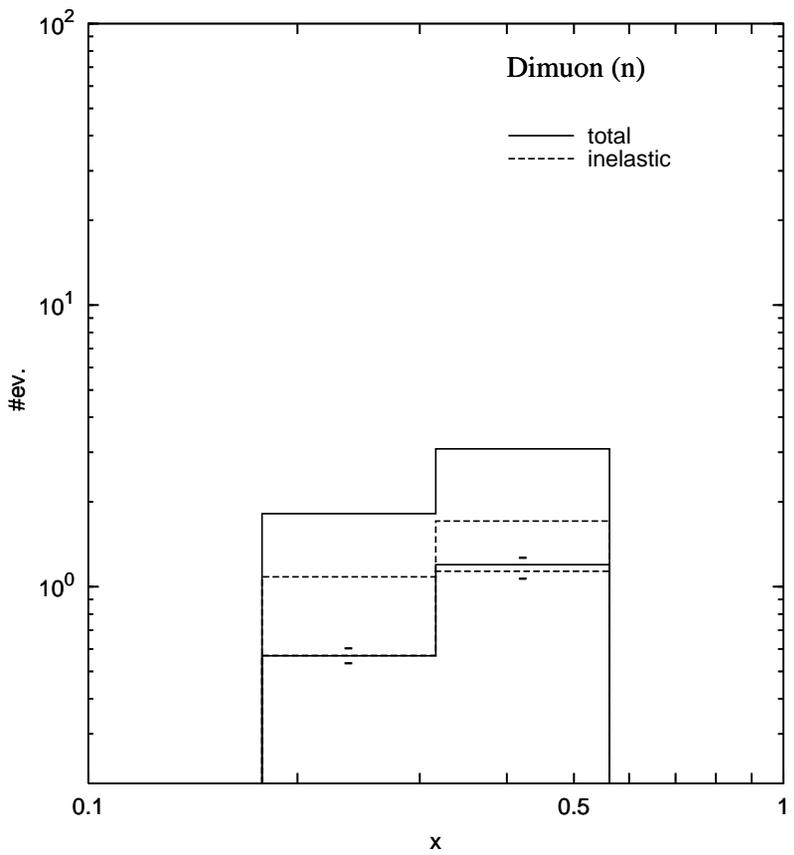, angle = 90, width = 11cm}

\vspace*{0.5cm}
\caption{
As in Fig.~\protect\ref{fig:fig5} but for 
dimuon production at HERMES using (un)polarized proton and
neutron targets. The lower solid and dashed curves refer to
a polarized nucleon target and the negative signs indicate that
the polarized cross sections are negative.}
\label{fig:fig9}
\end{figure}

\begin{figure}[t]
\centering
\epsfig{figure=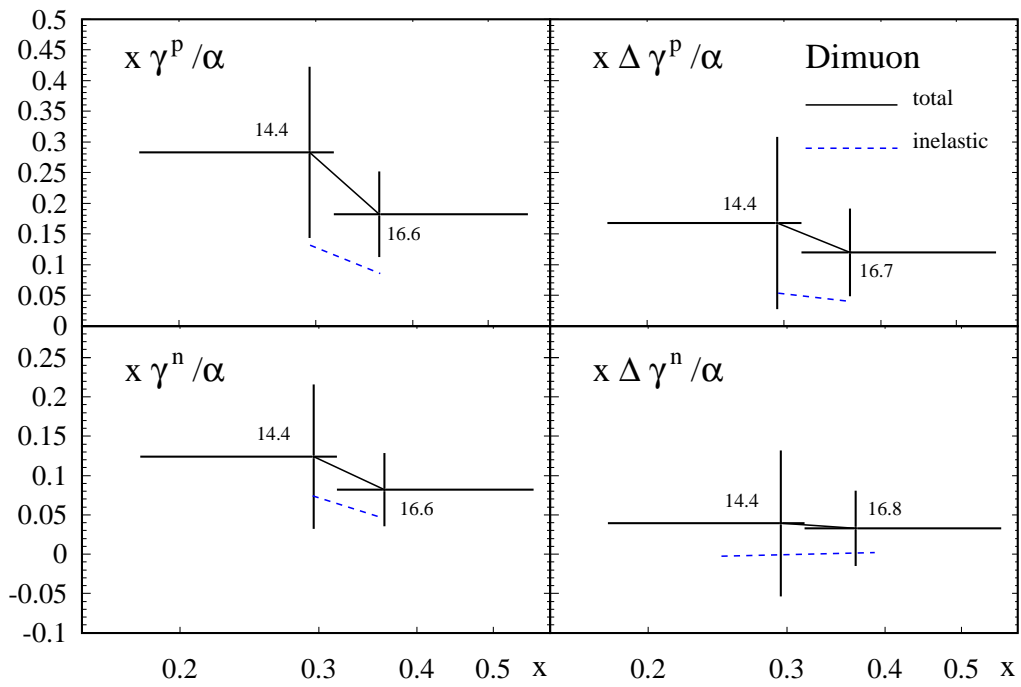, width = 14cm}

\vspace*{1cm}
\caption{
As in Fig.~\protect\ref{fig:fig6} but for 
dimuon production at HERMES for (un)polarized proton
and neutron targets. The statistical accuracy for the inelastic
contributions is similar to those shown for the total result, except
for the almost vanishing $\Delta \gamma^n_{\rm inel}$.}
\label{fig:fig10}
\end{figure}

For completeness, in Figs.~\ref{fig:fig9} and \ref{fig:fig10}
we also show the results for dimuon production at HERMES for
(un)polarized proton and neutron targets despite the fact that the
statistics will be far inferior to the Compton process.

The dimuon production can obviously proceed also via the genuine Drell-
Yan subprocess $q \bar{q}\to \mu^+ \mu^-$ where one of the (anti)quarks
resides in the resolved component of the photon emitted by the electron.
However, as already noted in \cite{Arteaga-Romero:1991wn}, this 
contribution is negligible as compared to the one due to the Bethe-
Heitler subprocess $\gamma\gamma \to \mu^+ \mu^-$. The unpolarized dimuon 
production rates at HERA where also studied in \cite{{Arteaga-Romero:1991wn},
{Bussey:1996vq}} utilizing, however, different prescriptions for
the photon content of the nucleon.\\
Exact expressions for the Bethe-Heitler contribution to the longitudinally
polarized $\gamma N \to \mu^+ \mu^- X$ process are presented in 
 \cite{Gehrmann:1997qh} but no estimates for the expected production
rates at, say, HERMES or COMPASS are given.

\section{Summary}
The analysis of the production rates of lepton-photon and muon pairs
at the colliding beam experiments at HERA and the fixed-target 
HERMES facility,
as evaluated in the leading order equivalent photon approximation,
demonstrates the feasibility of determining the polarized and
unpolarized equivalent photon distributions of the nucleon in the
available kinematical regions. The above mentioned production rates
can obviously be determined in a more accurate calculation along the
lines of \cite{Courau:1992ht}, involving the polarized 
and unpolarized
structure functions $g_{1,2}^N$ and $F_{1,2}^N$, repsectively, of
the nucleon.
The expected production rates are similar to those obtained in our
equivalent photon approximation (cf.\ Figs.\ 5.7 and 5.12 of
\cite{Lendermann:2002xx}).
It thus turns out that lepton-photon and muon pair production
at HERA and HERMES may provide an additional and independent
source of information concerning these structure functions.

\noindent{\large \bf Acknowledgments}\\
\noindent 
This work has been supported in part by the
 `Bundesministerium f\"ur Bildung und Forschung',
Berlin/Bonn.

\newpage
\bibliography{doth0216}

\begin{thebibliography}{13}
\expandafter\ifx\csname bibnamefont\endcsname\relax
  \def\bibnamefont#1{#1}\fi
\expandafter\ifx\csname bibfnamefont\endcsname\relax
  \def\bibfnamefont#1{#1}\fi
\expandafter\ifx\csname url\endcsname\relax
  \def\url#1{\texttt{#1}}\fi
\expandafter\ifx\csname urlprefix\endcsname\relax\def\urlprefix{URL }\fi
\expandafter\ifx\csname bibinfo\endcsname\relax \def\bibinfo#1#2{#2}\fi
\expandafter\ifx\csname eprint\endcsname\relax \def\eprint#1{#1}\fi

\bibitem{Gluck:2002fi}
\bibinfo{author}{\bibfnamefont{M.}~\bibnamefont{Gl{\"u}ck}},
  \bibinfo{author}{\bibfnamefont{C.}~\bibnamefont{Pisano}}, \bibnamefont{and}
  \bibinfo{author}{\bibfnamefont{E.}~\bibnamefont{Reya}},
  \bibinfo{journal}{Phys. Lett.} \textbf{\bibinfo{volume}{B540}}
  (\bibinfo{year}{2002}) \bibinfo{pages}{75}, \bibinfo{note}{{and references
  therein}}.

\bibitem{Kniehl:1991iv}
\bibinfo{author}{\bibfnamefont{B.~A.} \bibnamefont{Kniehl}},
  \bibinfo{journal}{Phys. Lett.} \textbf{\bibinfo{volume}{B254}}
  (\bibinfo{year}{1991}) \bibinfo{pages}{267}.

\bibitem{Gluck:1998xa}
\bibinfo{author}{\bibfnamefont{M.}~\bibnamefont{Gl{\"u}ck}},
  \bibinfo{author}{\bibfnamefont{E.}~\bibnamefont{Reya}}, \bibnamefont{and}
  \bibinfo{author}{\bibfnamefont{A.}~\bibnamefont{Vogt}},
  \bibinfo{journal}{Eur. Phys. J.} \textbf{\bibinfo{volume}{C5}}
  (\bibinfo{year}{1998}) \bibinfo{pages}{461}.

\bibitem{Gluck:2000dy}
\bibinfo{author}{\bibfnamefont{M.}~\bibnamefont{Gl{\"u}ck}},
  \bibinfo{author}{\bibfnamefont{E.}~\bibnamefont{Reya}},
  \bibinfo{author}{\bibfnamefont{M.}~\bibnamefont{Stratmann}},
  \bibnamefont{and}
  \bibinfo{author}{\bibfnamefont{W.}~\bibnamefont{Vogelsang}},
  \bibinfo{journal}{Phys. Rev.} \textbf{\bibinfo{volume}{D63}}
  (\bibinfo{year}{2001}) \bibinfo{pages}{094005}.

\bibitem{Courau:1992ht}
\bibinfo{author}{\bibfnamefont{A.}~\bibnamefont{Courau}} \bibnamefont{and}
  \bibinfo{author}{\bibfnamefont{P.}~\bibnamefont{Kessler}},
  \bibinfo{journal}{Phys. Rev.} \textbf{\bibinfo{volume}{D46}}
  (\bibinfo{year}{1992}) \bibinfo{pages}{117}.

\bibitem{DeRujula:1998yq}
\bibinfo{author}{\bibfnamefont{A.}~\bibnamefont{De~R{\'u}jula}}
  \bibnamefont{and}
  \bibinfo{author}{\bibfnamefont{W.}~\bibnamefont{Vogelsang}},
  \bibinfo{journal}{Phys. Lett.} \textbf{\bibinfo{volume}{B451}}
  (\bibinfo{year}{1999}) \bibinfo{pages}{437}. 

\bibitem{Drees:1994zx}
\bibinfo{author}{\bibfnamefont{M.}~\bibnamefont{Drees}},
  \bibinfo{author}{\bibfnamefont{R.~M.} \bibnamefont{Godbole}},
  \bibinfo{author}{\bibfnamefont{M.}~\bibnamefont{Nowakowski}},
  \bibnamefont{and} \bibinfo{author}{\bibfnamefont{S.~D.}
  \bibnamefont{Rindani}}, \bibinfo{journal}{Phys. Rev.}
  \textbf{\bibinfo{volume}{D50}} (\bibinfo{year}{1994}) \bibinfo{pages}{2335};

\bibitem{Ohnemus:1994qw}
\bibinfo{author}{\bibfnamefont{J.}~\bibnamefont{Ohnemus}},
  \bibinfo{author}{\bibfnamefont{T.~F.} \bibnamefont{Walsh}}, \bibnamefont{and}
  \bibinfo{author}{\bibfnamefont{P.~M.} \bibnamefont{Zerwas}},
  \bibinfo{journal}{Phys. Lett.} \textbf{\bibinfo{volume}{B328}}
  (\bibinfo{year}{1994}) \bibinfo{pages}{369}.

\bibitem{winp}
\bibinfo{note}{{Work in progress}}.

\bibitem{Brock:1995sz}
\bibinfo{author}{\bibfnamefont{G.}~\bibnamefont{Sterman}} \emph{et~al.},
  \bibinfo{collaboration}{CTEQ} Collaboration, \bibinfo{journal}{Rev. Mod.
  Phys.} \textbf{\bibinfo{volume}{67}} (\bibinfo{year}{1995}) 
  \bibinfo{pages}{157}.
  
\bibitem{Lendermann:2002xx}
\bibinfo{author}{\bibfnamefont{V.}~\bibnamefont{Lendermann}},
  \bibinfo{collaboration}{H1} Collaboration,
  \bibinfo{note}{{DESY-THESIS-2002-004}}.

\bibitem{Arteaga-Romero:1991wn}
\bibinfo{author}{\bibfnamefont{N.}~\bibnamefont{Arteaga-Romero}},
  \bibinfo{author}{\bibfnamefont{C.}~\bibnamefont{Carimalo}},
  \bibnamefont{and} \bibinfo{author}{\bibfnamefont{P.}~\bibnamefont{Kessler}},
  \bibinfo{journal}{Z. Phys.} \textbf{\bibinfo{volume}{C52}}
  (\bibinfo{year}{1991}) \bibinfo{pages}{289}.    

\bibitem{Bussey:1996vq}
\bibinfo{author}{\bibfnamefont{P.~J}~\bibnamefont{Bussey}},
  \bibinfo{author}{\bibfnamefont{B.}~\bibnamefont{Levtchenko}},
  \bibnamefont{and} \bibinfo{author}{\bibfnamefont{A.}~\bibnamefont{Shumilin}},   
  \bibinfo{note}{hep-ph/9609273}.

\bibitem{Gehrmann:1997qh}
\bibinfo{author}{\bibfnamefont{T.}~\bibnamefont{Gehrmann}}
  \bibnamefont{and} \bibinfo{author}{\bibfnamefont{M.}~\bibnamefont{Stratmann}},
  \bibinfo{journal}{Phys. Rev.} \textbf{\bibinfo{volume}{D56}}
  (\bibinfo{year}{1997}) \bibinfo{pages}{5839}.

\end{thebibliography}

\newpage
\pagestyle{empty}

\vspace*{3cm}

\newpage

\newpage

\newpage

\newpage

\newpage

\newpage

\newpage

\newpage

\end{document}